\documentclass[prl,aps,twocolumn, nopacs]{revtex4}
\usepackage[T1]{fontenc}
\usepackage[latin9]{inputenc}
\usepackage{amsmath}
\usepackage{graphicx}
\usepackage{times}
\usepackage[english]{babel}

\begin{document}

\title{Dynamical Entropy Production in Spiking Neuron Networks in the Balanced
State}

\pacs{87.19.lj, 87.10.-e, 05.45.-a, 05.10.-a}

\author{Michael Monteforte}

\email{monte@nld.ds.mpg.de}

\author{Fred Wolf}

\affiliation{Max Planck Institute for Dynamics and Self-Organization, G\"ottingen, Germany,\\
 Faculty of Physics, Georg-August-University G\"ottingen, G\"ottingen, Germany,\\
 Bernstein Center for Computational Neuroscience, G\"ottingen, Germany}

\begin{abstract}
We demonstrate deterministic extensive chaos in the dynamics of large
sparse networks of theta neurons in the balanced state. The analysis
is based on numerically exact calculations of the full spectrum of
Lyapunov exponents, the entropy production rate and the attractor
dimension. Extensive chaos is found in inhibitory networks and becomes
more intense when an excitatory population is included. We find a
strikingly high rate of entropy production that would limit information
representation in cortical spike patterns to the immediate stimulus
response. 
\end{abstract}

\maketitle

Neurons in the cerebral cortex fire action potentials (spikes) in
highly irregular, seemingly random sequences \cite{key-irregularity}.
Since neurons in isolation reliably respond to the repeated injection
of identical temporally varying inputs \cite{key-Mainen1995}, the
irregular activity in the cortex is not believed to result from a
randomness in the spike generating mechanism, but rather from strongly
fluctuating synaptic inputs \cite{key-Holt1996}. Several explanations
for the origin of such fluctuating inputs have been proposed \cite{key-ShadlenSoftky,key-VreeswijkSompolinsky}.
The prevailing explanation is a dynamic balance between excitatory
and inhibitory inputs, also known as the balanced state of cortical
networks. Such a balance in neuronal circuits has been demonstrated
experimentally in vitro and in vivo \cite{key-experiments}. Its statistical
characteristics have been studied theoretically in networks of excitatory
and inhibitory neurons \cite{key-theory,key-VreeswijkSompolinsky}
and in networks of only inhibitory neurons, where the recurrent inhibition
balances external excitatory currents \cite{key-LIF}. These studies
established that in sparsely connected networks with relatively strong
synapses the balanced state emerges robustly from the collective dynamics
of the network, without an external source of randomness.

The dynamical nature of the balanced state, however, remains controversial
and poorly understood. In a seminal paper, van Vreeswijk and Sompolinsky
developed a powerful mean field theory of the balanced state for excitatory-inhibitory
networks of binary neurons \cite{key-VreeswijkSompolinsky}. In their
framework, nearby trajectories diverged faster than exponentially,
demonstrating an extremely intense chaos with an infinite largest
Lyapunov exponent. More recently, studies of inhibitory networks of
leaky integrate and fire neurons reported stable chaos \cite{key-LIF}.
Stable chaos, originally discovered in coupled map lattices is characterized
by irregular but non-chaotic dynamics with a negative definite Lyapunov
spectrum \cite{key-stablechaos}.

Answering the question of whether the balanced state is chaotic is
of fundamental importance for understanding information representation
and processing in the cortex. If cortical networks were to generate
stable complex firing patterns, these might serve as a coding space
for storage and processing of, e.g., sensory information. If, alternatively,
cortical networks operate in a chaotic regime, information processing
would be intrinsically limited by the dynamical entropy production
that turns microscopic perturbations such as ion channel noise into
global firing patterns.

In this Letter, we present a comprehensive characterization of the
balanced state in networks of $N$ canonical type I neuronal oscillators,
called theta neurons \cite{key-thetanets}. The theta neuron model
explicitly describes dynamic action potential generation and is equivalent
to the normal form of a saddle node bifurcation on an invariant circle
\cite{key-thetaneuron}. In contrast to previously considered network
models that exhibit phase spaces of varying dimensionality \cite{key-Ashwin2005},
the phase space of our network model is a fixed $N$-torus. For these
networks, we performed numerically exact calculations of the full
spectrum of Lyapunov exponents, the rate of dynamical entropy production
and the attractor dimension, and analyzed the statistics of the first
Lyapunov vector. We found that both inhibitory networks and excitatory-inhibitory
networks exhibit deterministic extensive chaos. The rate of dynamical
entropy production in these networks was relatively high compared
to experimentally measured information content of spike patterns in
sensory cortices.

We studied large sparse networks of $N$ theta neurons arranged on
directed Erd\"os-R\'enyi random graphs of fixed mean indegree $K$.
The theta neuron model is a phase representation of the quadratic
integrate and fire model \cite{key-thetaneuron}. Neurons are described
by phases $\theta_{i}\in[-\pi,\pi]$, with spikes defined to occur
when $\theta_{i}$ crosses $\pi$. The governing equation is 
\begin{equation}
    \tau_{\textrm{m}}\frac{\textrm{d}\theta_{i}}{\textrm{d}t}=(1-\cos\theta_{i})+I_{i}(t)(1+\cos\theta_{i}),\label{eq:thetaneuron}
\end{equation}
with the membrane time constant $\tau_{\textrm{m}}$ and the synaptic
input current 
\begin{equation}
    I_{i}(t)=\sqrt{K}I_{i}^{\textrm{ext}}+2\tau_{\textrm{m}}\sum_{j\in\textrm{pre}(i)}\frac{J_{ij}}{\sqrt{K}}\,\delta(\theta_{j}(t)-\pi).\label{eq:current}
\end{equation}
All neurons $i=1\dots N$ receive constant external excitatory currents
$\sqrt{K}I_{i}^{\textrm{ext}}$. These were identical for all neurons
in the excitatory (E) and inhibitory (I) population, respectively,
and chosen to obtain a target average firing rate $\bar{\nu}_{E}=\bar{\nu}_{I}=\bar{\nu}$.
At spike events of a presynaptic neuron $j\in\textrm{pre }(i)$, neuron
$i$ received non-delayed $\delta$-pulses of strength $J_{ij}/\sqrt{K}$,
leading to step-like changes in the neuron's state. In inhibitory
networks, every neuron was coupled to on average $K$ presynaptic
neurons with $J_{II}=-J_{0}$. In excitatory-inhibitory networks,
the inhibitory neurons received inputs from on average $K$ inhibitory
neurons with $J_{II}=-\sqrt{1-\varepsilon^{2}}J_{0}$ and $K$ excitatory
neurons with $J_{IE}=\varepsilon J_{0}$. The excitatory neurons received
inputs from on average $K$ excitatory neurons with $J_{EE}=0.9\varepsilon J_{0}$
and $K$ inhibitory neurons with $J_{EI}=-\sqrt{1-(0.9\varepsilon)^{2}}J_{0}$.
At $\varepsilon=0$, all excitatory neurons are passive, since they
only receive inputs from the inhibitory neurons without providing
feedback. Increasing $\varepsilon$ activates the excitatory feedback
loops, such that the magnitudes of input fluctuations $\sigma^{2}=J_{0}^{2}\bar{\nu}$
are preserved and identical to those in inhibitory networks \cite{key-Supplement}.

From the analytical solutions of Eq.~\eqref{eq:thetaneuron} with
\eqref{eq:current} we obtained an N-dimensional map of the neurons'
phases between successive spike times in the network $\{t_{s}\}$
\cite{key-Supplement}. This map was used for event-based, numerically
exact simulations of the network dynamics and yields the single spike
Jacobian matrix $D(t_{s})=\frac{\partial\vec{\theta}(t_{s})}{\partial\vec{\theta}(t_{s-1})}$:
\begin{equation}
    D_{ij}(t_{s})=
    \begin{cases}
        d_{i^{*}}(t_{s}) & \textrm{for }i=j=i^{*}\\
        \sqrt{\frac{I_{i^{*}}^{\textrm{ext}}}{I_{j^{*}}^{\textrm{ext}}}}\big(1-d_{i^{*}}(t_{s})\big) & \textrm{for }i=i^{*},\, j=j^{*}\\
        \delta_{ij} & \textrm{otherwise},
    \end{cases}
    \label{eq:jacobian}
\end{equation}
where $j^{*}$ denotes the spiking neuron at time $t_{s}$, $i^{*}\in\textrm{post (}j^{*})$
the spike receiving neurons, $\delta_{ij}$ is the Kronecker symbol
and 
\begin{equation}
    d_{i^{*}}(t_{s})=\frac{\big(\!\tan(\theta_{i^{*}}(t_{s}^{-})/2)\big)^{2}+\sqrt{K}I_{i^{*}}^{\textrm{ext}}}{\big(\!\tan(\theta_{i^{*}}(t_{s}^{-})/2)+J_{i^{*}j^{*}}/\sqrt{K}\,\big)^{2}+\sqrt{K}I_{i^{*}}^{\textrm{ext}}},
\end{equation}
where $\theta_{i^{*}}(t_{s}^{-})$ denotes the phase directly before
spike reception \cite{key-Supplement}.

The exact Jacobians \eqref{eq:jacobian} were used to numerically
calculate all Lyapunov exponents in the standard Gram--Schmidt reorthogonalization
procedure \cite{key-Supplement,key-Benettin1980}. From the Lyapunov
exponents $\lambda_{1}\ge\ldots\ge\lambda_{N}$, we derived the Kolmogorov--Sinai
entropy production (Pesin identity): $H=\sum_{\lambda_{i}>0}\lambda_{i}$
and the attractor dimension (Kaplan--Yorke conjecture): $D=d+S_{d}/|\lambda_{d+1}|$
(with $d$ the largest integer such that $S_{d}=\sum_{i=1}^{d}\lambda_{i}\ge0$)
\cite{key-EckmannRuelle}. All calculations were repeated with 10
different initial conditions and network realizations. In all Figures
below, standard errors are smaller than the symbol sizes.

\begin{figure}
\includegraphics[clip,width=1\columnwidth]{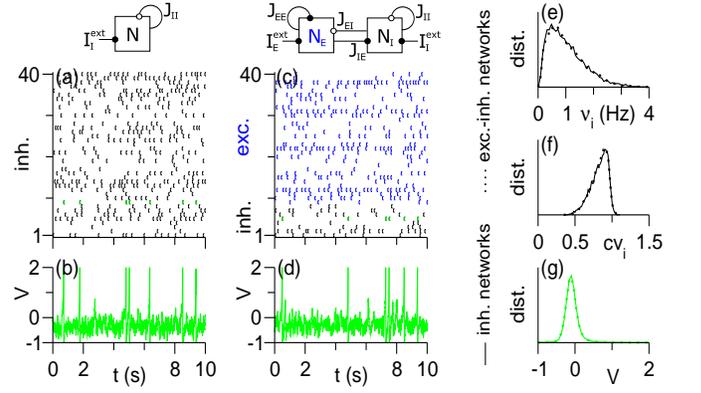} 

\caption{\label{fig:fig1}(color online) Characteristics of the balanced state
in inhibitory and excitatory-inhibitory networks of theta neurons:
(a,c) Spike patterns of 40 random neurons, (b,d) Voltage traces of
one random neuron, (e) Firing rate distributions, (f) Coefficient
of variation distributions and (g) Stationary voltage distributions,
(parameters: $N=10000$, $K=100$, $\bar{\nu}=1\,\textrm{Hz}$, $J_{0}=1$,
$\tau_{\textrm{m}}=10\textrm{ ms}$, $r=0.9$, $\varepsilon=0.3$,
$N_{E}=4N_{I}$)}

\end{figure}

The characteristics of inhibitory networks and excitatory-inhibitory
networks in the balanced state are depicted in Fig.~\ref{fig:fig1}.
Representative spike patterns and voltage traces $V(t)=\tan(\theta(t)/2)$
illustrate the irregular, asynchronous firing and strong membrane
potential fluctuations. In agreement with recent biological observations,
the firing rate distributions were broad \cite{key-lognormal}, indicating
a substantial heterogeneity in the networks. Distributions of firing
rates and coefficients of variation were identical in both types of
networks.

\begin{figure*}
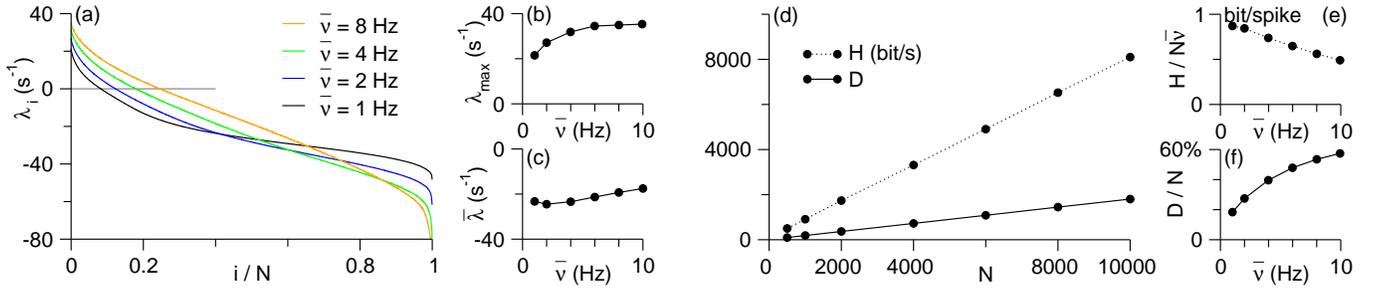

\includegraphics[clip,width=1\columnwidth]{fig2a} \hfill{}\includegraphics[clip,width=1\columnwidth]{fig2b} 

\caption{\label{fig:fig2}(color online) Extensive deterministic chaos in balanced
inhibitory networks: (a) Full Lyapunov spectra for different average
firing rates $\bar{\nu}$, collapsed for different network sizes,
(b) Maximal Lyapunov exponent, (c) Mean Lyapunov exponent, (d) Entropy
production rate $H$ and attractor dimension $D$ for various network
sizes $N$, (e) Average entropy production per spike per neuron, (f)
Attractor dimension density (parameters: $N=2000$, $K=100$, $\bar{\nu}=1\,\textrm{Hz}$,
$J_{0}=1$, $\tau_{\textrm{m}}=10\textrm{ ms}$)}

\end{figure*}

We first characterized the dynamics of inhibitory networks in the
balanced state (Fig.~\ref{fig:fig2}). These networks exhibited deterministic
extensive chaos, characterized by finite positive maximal Lyapunov
exponents and network size invariant Lyapunov spectra \cite{key-Supplement}.
As a result, the attractor dimension and entropy production increased
linearly with the number of neurons in the networks. The range of
the Lyapunov spectra increased with the average firing rate, such
that the maximal Lyapunov exponent, entropy production and attractor
dimension grew for larger firing rates. The chaotic attractors had
a large dimensionality of the order of the phase space dimension $N$.
The average entropy production (information loss) was of $\mathcal{O}(1)$
bit per spike per neuron. We remark that this is surprisingly high
compared to the estimated information content of $\mathcal{O}(1)$
bit per spike in primary sensory cortices \cite{key-Borst1999,key-Panzeri}.

\begin{figure}
\includegraphics[clip,width=1\columnwidth]{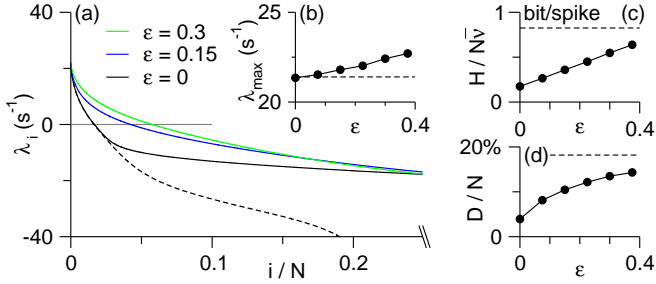} 

\caption{\label{fig:fig3} (color online) Chaotic dynamics in balanced excitatory-inhibitory
networks at different excitatory feedback loop activations $\varepsilon$:
(a) Lyapunov spectra, (b) Maximal Lyapunov exponent, (c) Entropy production
per spike per neuron, (d) Attractor dimension density (dashed lines:
isolated inhibitory networks with $N=N_{I}$, parameters: $N_{\textrm{I}}=2000$,
$N_{\textrm{E}}=8000$, $K=100$, $\bar{\nu}=1\,\textrm{Hz}$, $J_{0}=1$,
$\tau_{\textrm{m}}=10\textrm{ ms}$)}

\end{figure}

Adding an excitatory population to the networks increased the intensity
of balanced state chaos (Fig.~\ref{fig:fig3}), while the magnitude
of input fluctuations and the basic firing statistics were preserved.
When the excitatory neurons are passive ($\varepsilon=0$), their
dynamics is fully controlled by the inhibitory neurons. The positive
part of the Lyapunov spectrum was then equivalent to that of an isolated
inhibitory network of reduced size $N=N_{\textrm{I}}$ (dashed line).
Hence, the maximal Lyapunov exponent was identical, whereas the entropy
production per spike per neuron and the attractor dimension density
were reduced accordingly by one fifth. Upon activation of the excitatory
feedback loops ($\varepsilon>0$) the entire Lyapunov spectrum changed.
The maximal Lyapunov exponent, attractor dimension and entropy production
increased with $\varepsilon$, demonstrating an intensification of
the chaos. 

\begin{figure}
\includegraphics[clip,width=1\columnwidth]{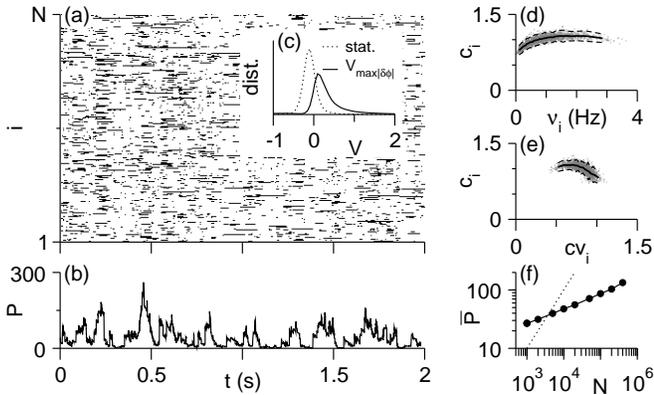} 

\caption{\label{fig:fig4}Temporal network chaos in balanced inhibitory networks:
(a) First Lyapunov vector $\vec{\delta\phi}(t)$ (marked black if
$|\delta\phi_{i}(t)|>1/\sqrt{N}$), (b) Participation ratio $P$(t)
of first Lyapunov vector, (c) Voltage distribution of neurons with
largest Lyapunov vector elements (dotted line: stationary distribution)
(d,e) Neurons' chaos indices $c_{i}$ versus firing rates $\nu_{i}$
and coefficients of variation $cv_{i}$ (straight line: mean, dashed
lines: $\pm2\textrm{ std}$), (f) Average participation ratio $\bar{P}$
vs.~network size $N$ (dotted line: guide to the eye for $\bar{P}\sim N$),
(parameters: $N=2000$, $K=100$, $\bar{\nu}=1\,\textrm{Hz}$, $J_{0}=1$,
$\tau_{\textrm{m}}=10\textrm{ ms}$)}

\end{figure}

Are all neurons involved equally and at all times in the chaotic dynamics
despite their substantial firing heterogeneities? To answer this question,
we studied the first Lyapunov vector $\vec{\delta\phi}(t)$ ($\sum_{i=1}^{N}\delta\phi_{i}(t)^{2}=1$).
At each point in time, $\vec{\delta\phi}$ aligns with the direction
in which initial perturbations exponentially grow with maximal asymptotic
rate $\lambda_{\textrm{max}}$. The Lyapunov vector was dominated
by relatively small subsets of neurons that changed over time (Fig.~\ref{fig:fig4}).
We analyzed the composition of these groups with two quantities: the
participation ratio $P(t)=1/\sum_{i=1}^{N}\delta\phi_{i}(t)^{4}$
quantifying the size of a group at time $t$ \cite{key-Cross1993},
and the chaos index $c_{i}=\sqrt{N\langle\delta\phi_{i}(t)^{2}\rangle_{t}}$
quantifying the time-averaged participation of individual neurons.
The participation ratio exhibited substantial fluctuations indicating
strongly varying group sizes. For the considered rates $1\,\textrm{Hz}\le\bar{\nu}\le10\,\textrm{Hz}$,
the time-averaged participation ratio $\bar{P}=\langle P(t)\rangle_{t}$
obeyed a sublinear scaling $\bar{P}\sim N^{\alpha}$, with $0.25\le\alpha\le0.5$
\cite{key-Supplement}. Thus, although the average group size grew
with network size, the fraction of most unstable neurons decreased
algebraically as $N^{\alpha-1}$. The chaos indices were essentially
independent of the neurons' firing rates and cv values. This means
that statistical firing irregularity is unrelated to chaoticity. Thus,
although at any time only a small group of neurons constitutes the
most unstable direction, over time all neurons participated almost
equally in the chaotic dynamics. 

To understand when the neurons were the most sensitive to state perturbations,
we examined the voltage distribution of neurons when exhibiting the
largest Lyapunov vector element. This distribution compared to the
stationary distribution was shifted towards the spike generating instability,
marking it as an important source of instability of the collective
dynamics (Fig.~\ref{fig:fig4}(c)).

The chaotic dynamics of balanced networks described here shares many
features with classical examples of spatio-temporal chaos \cite{key-Cross1993}.
Balanced networks exhibit a size-invariant Lyapunov spectrum and extensive
entropy production rate and attractor dimension. The chaotic degrees
of freedom are confined to small and varying groups of elements. The
type of this localization, however, differs from classical examples
of spatio-temporal chaos. While classically, the participation ratio
is constant for large $N$, we here found an algebraic sublinear dependence.
This behavior apparently reflects the non-local graph structure of
the systems and might thus be termed temporal network chaos. In fact,
chaos in balanced networks is induced by the complex network connectivity:
neither the isolated units nor a fully connected network of theta
neurons exhibit chaotic dynamics. Extensive chaos is expected in spatially
extended systems, that can be decomposed into weakly interacting subsystems,
whose number grows linearly with system size \cite{key-Ruelle1982}.
As this is not fulfilled for network dynamical systems on random graphs,
extensive chaos in our networks is not a trivial property. Globally
coupled networks for instance can in fact exhibit non-extensive chaos
\cite{key-Chate}. The extensivity found for balanced chaos likely
reflects the lack of correlation between neurons in this state. Even
in densely connected random networks, pairwise correlations are expected
to vanish in the large system limit \cite{key-async}. Then, the invariant
measure will factorize and the attractor becomes $\mathcal{O}(N)$-dimensional
in the large system limit. It should be noted that the studied networks
may be viewed as subunits of cortical networks that can be approximated
with random connectivity. In spatially extended networks, composed
of such subunits, extensivity may be expected for standard reasons.

Our results reveal that the dynamics of balanced networks strongly
depends on the single neuron dynamics. While we found that theta neurons
in the balanced state exhibit conventional extensive chaos, previous
studies of networks of leaky integrate and fire neurons demonstrated
stable chaos \cite{key-LIF}. Opposed to the leaky integrate and fire
model, the theta neuron model incorporates the dynamic instability
underlying spike initiation. The membrane potential distribution of
the neurons\textbf{ }when most sensitive to perturbations is indeed
shifted towards this instability. Our results thus indicate, that
this feature of single neuron dynamics renders the collective dynamics
of balanced networks robustly chaotic.

The strikingly large entropy production we found raises questions
about the neural code in cortical networks. If the entropy production
is of the same order of magnitude as sensory information carried by
spikes, sensory information would be hard to maintain in spike patterns
beyond the immediate stimulus response. In fact, a study of spike
timing in barrel cortex showed that information about whisker deflections
is encoded to 83\% in the first spike after the stimulus and reduced
in each successive spike by roughly two thirds \cite{key-Panzeri}.
Since the theta neuron model is the canonical form of type I excitability,
our results are expected to be representative of a wide range of single
neuron dynamics. The reliable generation of long and precisely timed
spike sequences by cortical networks would then be unlikely. 

The fat chaotic attractors in balanced networks described here might
serve as a rich repertoire of complex states the network is rapidly
plowing through. Our study lays the foundation for future research
on dynamic stabilization mechanisms for unstable states or reliable
computations on short time scales on existing stable directions.

We thank E.~Bodenschatz, P.~Cvitanovi\'c, T.~Geisel, M.~Gutnick,
H.~Sompolinsky, M.~Timme and C.~van Vreeswijk for fruitful discussions.
M.~M.~acknowledges the hospitality of J.~Golowasch and F.~Nadim
at Rutgers University. This work was supported by BMBF (01GQ07113)
and GIF (906-17.1/2006).

Note added in proof: Our findings of a strongly chaotic dynamics is
consistent with novel experimental evidence of a high sensitivity
to perturbations in cortical networks in vivo \cite{key-Latham}.

\end{document}